\documentclass[runningheads]{llncs}
\usepackage[ruled,vlined, linesnumbered]{algorithm2e}
\usepackage{subfig}
\newcommand{\quotes}[1]{``#1''}
\usepackage{tabularx}
\usepackage{graphicx}
\usepackage{adjustbox}
\usepackage{longtable}
\usepackage{tablefootnote}
\usepackage{color,soul}
\usepackage{booktabs}
\usepackage{svg}
\usepackage{amssymb}
\usepackage{amsmath}

\AtBeginDocument{%
  \providecommand\BibTeX{{%
    \normalfont B\kern-0.5em{\scshape i\kern-0.25em b}\kern-0.8em\TeX}}}
    
\begin{document}
\title{FiLMing Multimodal Sarcasm Detection with Attention} \author{Sundesh Gupta\inst{1} \and
Aditya	Shah \inst{1} \and
Miten	Shah \inst{1} \and
Laribok	Syiemlieh \inst{2} \and
Chandresh	Maurya \inst{1}}

\institute{ Indian Institute of technology Indore, India\\ \email{\{cse180001057,cse180001049,chanderesh@iiti.ac.in\}}\\ \email{adishah3103@gmail.com}
\and 
NIT Meghalaya, India \\\email{lariboksyiemlieh@gmail.com}}

%
%

\maketitle              
\begin{abstract}
 Sarcasm detection identifies natural language expressions whose intended meaning is different from what is implied by its surface meaning. It finds applications in many NLP tasks such as opinion mining, sentiment analysis, etc. Today, social media has given rise to an abundant amount of multimodal data where users express their opinions through text and images. Our paper aims to leverage multimodal data to improve the performance of the existing systems for sarcasm detection. So far, various approaches have been proposed that uses text and image modality and a fusion of both. We propose a novel architecture that uses the RoBERTa model with a co-attention layer on top to incorporate context incongruity between input text and image attributes.
Further, we integrate feature-wise affine transformation by conditioning the input image through FiLMed ResNet blocks with the textual features using the GRU network to capture the multimodal information. The output from both the models and the CLS token from RoBERTa is concatenated and used for the final prediction. Our results demonstrate that our proposed model outperforms the existing state-of-the-art method by \textbf{6.14\% F1 score} on the public Twitter multimodal sarcasm detection dataset.
\end{abstract}

\section{Introduction}
According to wiki, \emph{sarcasm is a sharp, bitter, or cutting expression or remark; a bitter gibe or taunt}. Thus, sarcasm is defined as a sharp remark whose intended meaning is different from what it looks like. For example, \quotes{I am not insulting you. I am describing you}, could mean that the speaker is insulting the audience, but the receiver does not get it. Sarcasm does not necessarily involve irony, but more often than not, sarcasm is used as an ironic remark \footnote{www.thefreedictionary.com.}. Sarcasm usually involves ambivalence and is difficult to comprehend. It requires a certain degree of intelligence quotient (IQ) and age to deliver sarcasm or understand it \cite{lumen}. Sarcasm detection is an important task in many natural language understanding tasks such as opinion mining, dialogue systems, customer support, online harassment detection, to name a few. In particular, psychologists use the ability to understand sarcasm as a tool to distinguish among different types of neuro-degenerative diseases \cite{sarcasm}.

A plethora of works automatically detects sarcasm in unimodal data, using either text or images. Over the years, online media and chatbots have given rise to multimodal data such as images, texts, video, and audio. In the present work, we consider sarcasm detection using image and text modality only. The scope of our work encompasses where the text and image have opposite meaning. That means, we do not consider the case where only the text is sarcastic or only the image is sarcastic and leave it for future studies. 

\begin{figure}
\centering
    \subfloat[lovely, clean, pleasant train home]{
    \includegraphics[width=4cm, height=3cm]{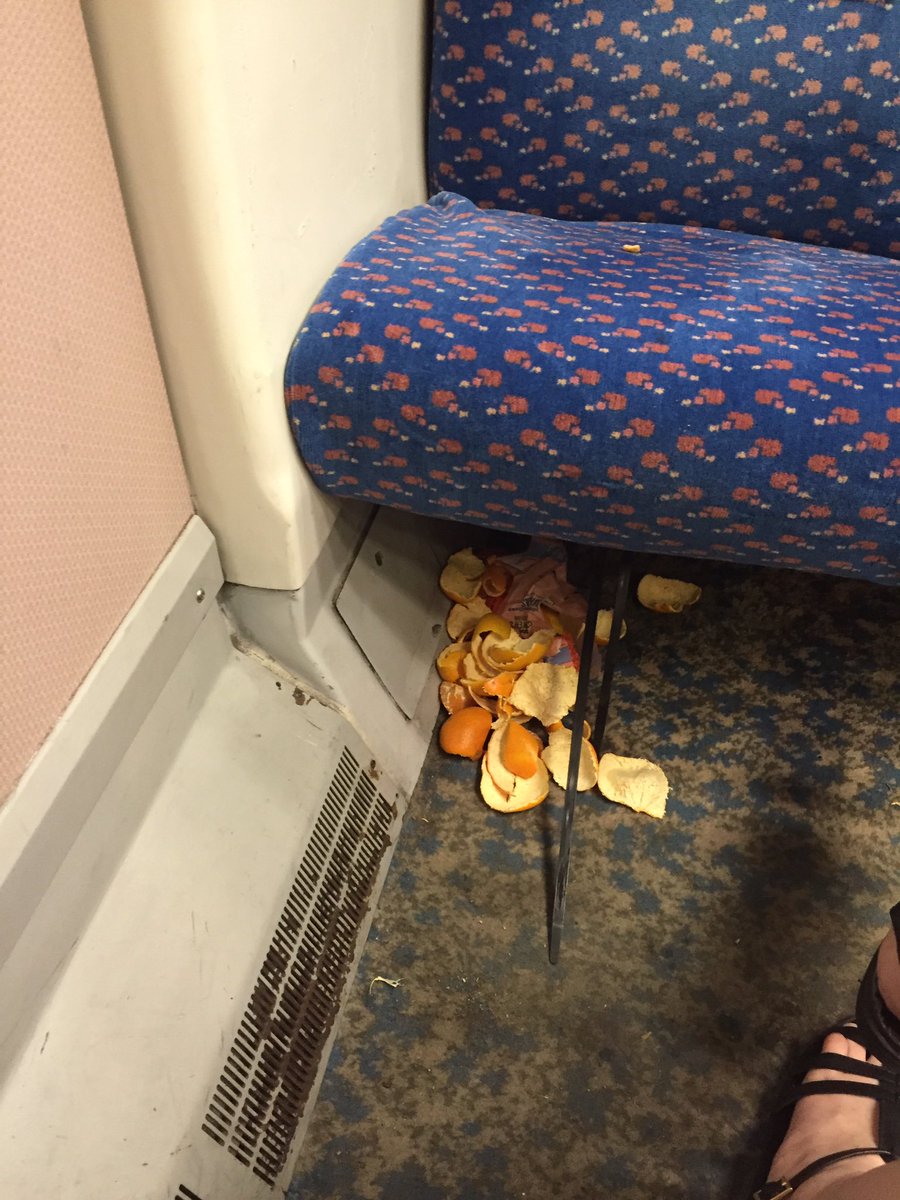}}
    \hspace{2em}
    \subfloat[well that looks appetising \#ubereats]{
    \includegraphics[width=4cm, height=3.2cm]{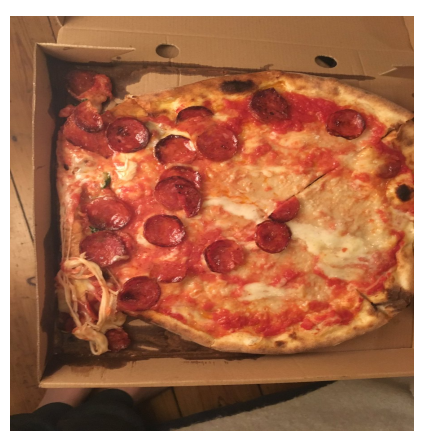}}
    \caption{Examples from the Twitter data showing text modality alone is insufficient for sarcasm detection.}
    \label{multisarcasm}
 \vspace{-6mm}
\end{figure}

One of the examples of sarcasm in multimodal data is presented in fig. \ref{multisarcasm}. Detecting sarcasm in multimodal data can be more challenging as compared to unimodal data simply because what the text says is the opposite of what the image implies, for example, in fig. \ref{multisarcasm}(a), the text says \quotes{lovely, clean, pleasant train home} whereas the associated image implies the opposite. Similarly, in fig. \ref{multisarcasm}(b), the textual description and image semantics are alluding to opposite meanings. Such a phenomena is called \emph{incongruity} \cite{joshi2015harnessing,xiong2019sarcasm,riloff2013sarcasm} and has been leveraged to tackle multimodal sarcasm detection  \cite{sangwan2020didn,xu2020reasoning,wang2020building,schifanella2016detecting}.

Following previous multimodal sarcasm detection approaches, we propose a deep learning-based architecture that takes as input text and image modality. We compute the inter-modality incongruity in two ways. Firstly, the image attribute is extracted using ResNet \cite{he2016deep}, and the co-attention matrix is calculated. This operation captures the inter-modality incongruity between the text and image attribute. Secondly, the inter-modality incongruity is computed between the text and image features. The two incongruity representations are fused together and used for classifying the sentences into sarcastic or non-sarcastic. Concretely, we make the following contributions: (1)  A novel deep learning-based architecture is proposed that captures the inter-modality incongruity between the image and text, (2)  Empirical demonstration shows that we are able to boost the F1 score by 6.14\% of the current SOTA for multimodal sarcasm detection on the benchmark Twitter dataset.
\section{Related Works}
Various methods have been proposed in the literature for sarcasm detection in unimodal and multimodal data. In this section, we delineate works that tackle sarcasm detection in multimodal data using image and text modality only. Using audio or video is out of scope of this work and hence not presented here.
One of the first works to utilize multimodal data for sarcasm detection is \cite{schifanella2016detecting} which presents two approaches. The first one exploits visual semantics trained on an external dataset and concatenates the semantic features with the textual features. The second method adopts a visual neural network initialized with parameters trained on ImageNet for multimodal sarcastic posts. Cognitive features are extracted using gaze/eye-movement data, and CNN is used to encode them for feature representation in \cite{mishra2017learning}. The work in \cite{cai2019multi} extracts visual features and visual attributes from images using ResNet and builds a hierarchical fusion model to detect sarcasm.
Along the same lines, the recurrent network model in \cite{sangwan2020didn} proposes the idea of a gating mechanism to leak information from one modality to the other and achieves superior performance on Twitter benchmark dataset for sarcasm detection. The authors of \cite{wang2020building} use pre-trained BERT and ResNet models to encode text and image data and connect the two using a gate called a bridge. Further, they also propose a 2D-Intra-Attention layer to extract the relationship between the text and image.

The multimodal work that closely matches with our work is \cite{pan2020modeling}. In that, the author proposes a BERT-based architecture for modeling intra- and inter-modality incongruity. Self-attention is used to model inter-modality incongruity, whereas co-attention is used to model intra-modality incongruity. Intra-modality incongruity is also used in \cite{xiong2019sarcasm} for sarcasm detection. Contrary to this work, we model inter-modality incongruity between the \emph{text} and \emph{visual attributes} (The work in \cite{pan2020modeling} uses \emph{text} and \emph{visual features}) in {\bf two ways}. The first uses visual attributes and text, and the latter uses visual features and text. Note that visual features and visual attributes are two different entities. The former is a low-level representation of the image, whereas the latter is high-level description of the image such as what objects are present in the image? The second major difference is that we use feature-wise linear transformation \cite{perez2018film} to compute FiLM parameters using the text data and inject the FiLM layers (see \cite{perez2018film} for more details) in between the ResNet layers. The FiLMed ResNet outputs the visual features, which acts like the inter-modality incongruity. More details are given in the next section.

\section{Methodology}



\subsection{Proposed Model}
\begin{figure}
\centering
\includegraphics[width=11.5cm, height=8.5cm]{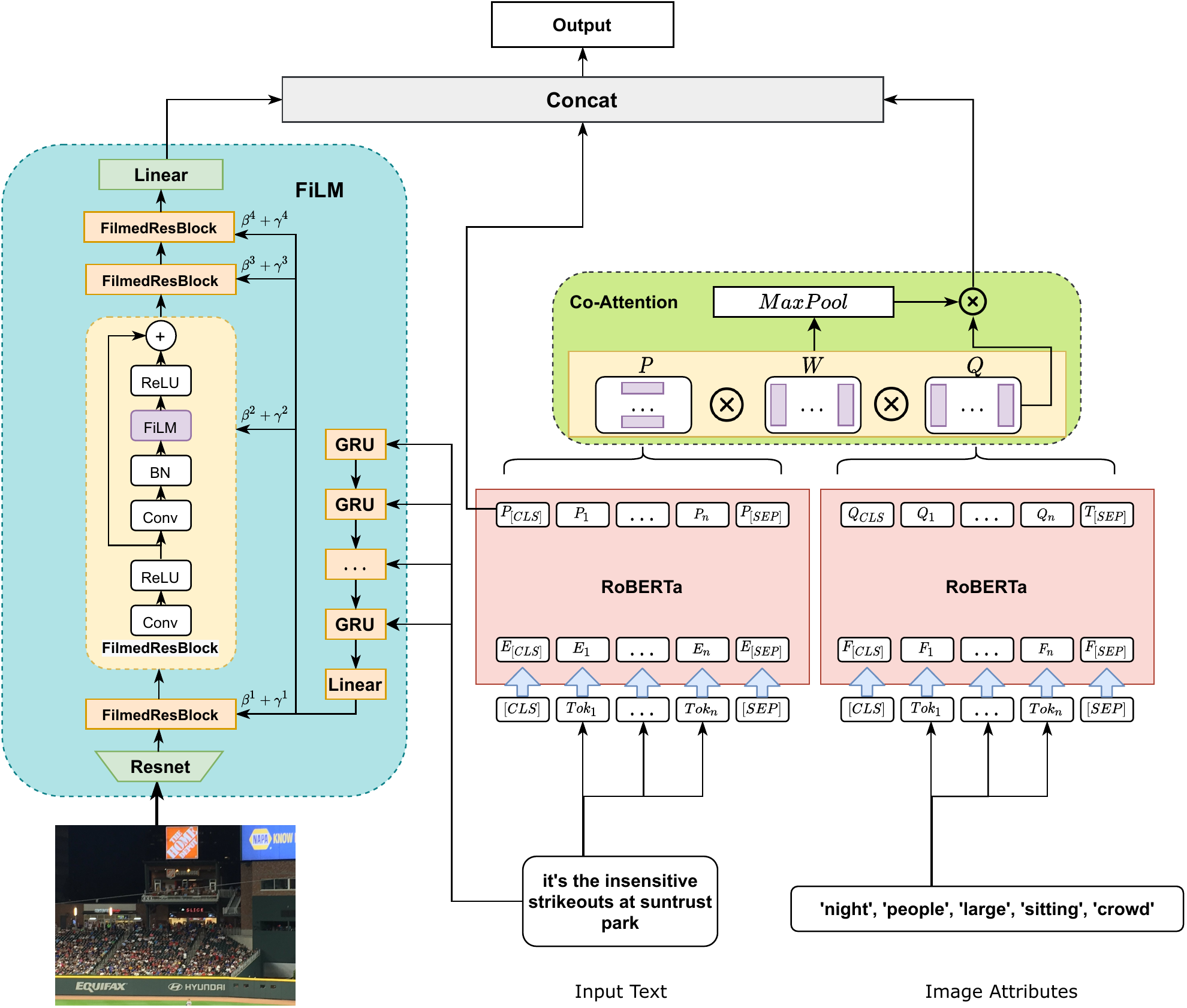}
\caption{Overview of our proposed model}
\label{architecture}
\end{figure}

We propose a novel multimodal architecture using Robustly optimized Bidirectional Encoder Representation from Transformer (RoBERTa) \cite{liu2019roberta} for detecting sarcasm. Figure \ref{architecture} gives an overview of the model. The proposed model mainly consists of the text and image attribute representation, image representation conditioned on text (FiLM), multimodal incongruity through the co-attention mechanism, and final concatenation with [CLS] token.  The description of each component is elucidated in the next section.

\subsubsection{Image, text and image attribute representation}

For representing text, we  consider  them  as a sequence of words $E = \{[CLS], E_1, E_2, \ldots, E_n, [SEP]\}$, here $E_i \in \mathbb{R}^d$ is the sum of token, segment and position embeddings, $n$ denotes maximum length of the input text, and $d$ is the embedding size. For extracting features $P \in \mathbb{R}^{N \times d}$, we use the RoBERTa model and consider the output of the first encoder layer of RoBERTa as the representation of the text. Here, $N$ depicts the length of set $E$ and $d$ is the hidden size of RoBERTa. Similarly, for the representation of image attributes, we have $F = \{[CLS], F_1, F_2, \ldots, F_m, [SEP]\}$ as the sum of the token, segment, and position embeddings, and its features will be represented by $Q \in \mathbb{R}^{M \times d}$ which is the output of the last layer of RoBERTa. Here, $M$ is the length of set F.

\subsubsection{Inter-modal incongruity between visual and text representation}: Since input text plays an important role in detecting sarcasm, we capture the image information based on the textual features. Inspired by the work of \cite{perez2018film}, we apply the feature-wise affine transformation on the image by conditioning it on the input text. The image features are extracted using the pre-trained ResNet-50. Further, we use Gated Recurrent Unit (GRU) network  \cite{DBLP:journals/corr/ChungGCB14} to process the text, which takes 100-dimensional learned GloVe \cite{pennington-etal-2014-glove} word embeddings as the input. The final layer of the GRU network outputs FiLM parameters $(\gamma_i^n$, $\beta_i^n)$ for $n^{th}$ FiLMed residual block. The  $\gamma_i$ and $\beta_i$ are the output of the functions $g$ and $h$, which are learned by the FiLM for the input $x_i$:

\begin{equation}
    \gamma_{i,c} = g_c(x_i)  \hspace{1cm}   \beta_{i,c} = h_c(x_i), 
\end{equation}

where $g$, $h$ are arbitrary functions.  

In our experiment, we use 4 FiLMed residual blocks with a linear layer attached on top which outputs the final output $Q_{film} \in R^{1024}$. Using the FiLM parameters, FiLM layers are inserted within each residual block to condition the visual pipeline. Mathematically, the parameters $\gamma$ and $\beta$ perform the feature-wise affine transformation on the image feature maps extracted by ResNet
\begin{equation}
    FiLM(F_{i,c}) = \gamma_{i,c} * F_{i,c} + \beta_{i,c}
\end{equation} 
Here, $F_{i,c}$ corresponds to the $i^{th}$ input's $c^{th}$ image feature map.
Doing so, we aim to extract visual features akin to text meaning. On the other hand, conditioning the image representation on text representation captures the inter-modal incongruity.

\subsubsection{Inter-modal incongruity between visual attributes and text representation} 

To model the contradiction between the input-text and image attribute, we use a co-attention mechanism motivated by \cite{lu2016coatt}. The co-attention input is the RoBERTa model's output of input text and high-level image representation, i.e., image attributes. The co-attention mechanism incorporates the incongruity between the text and image modality. Formally, we first calculate the affinity matrix \(C\) using bi-linear transformation $W$ to capture the interaction between the input text and the image attributes.

\begin{equation}
    C = tanh(PWQ^T)
\end{equation}

where $P \in R^{N \times d} $ represent input-text features, $Q \in R^{M \times d} $ represent image attribute features, $W \in R^{d \times d} $ is a learnable parameter consisting of weights. $N$ and $M$ denote the maximum size of input-text features and image attributes features, respectively, and $d$ denotes the hidden-size of RoBERTa. The affinity matrix $C$ transforms the text attention space to image attribute attention space. The attention weight $\alpha \in R^M $ is then calculated using 2D max-pooling operation over affinity matrix C using a kernel of size ($N \times 1$). Intuitively, $\alpha$ calculates the attention weights over each word in the text, which has been transformed to image attribute attention space.

\begin{equation}
    \alpha = MaxPool(C)
\end{equation}

Finally, the image attribute attention matrix, $Q_{att} \in R^d$, which captures the contradiction between text and high-level features of image is calculated as:

\begin{equation}
    Q_{att} = \alpha Q
\end{equation}

\subsubsection{Final Fusion} 


Summing up together, we take the $Q_{film} \in \mathbb{R}^{1024}$ output from the FiLM and $Q_{att}$ from co-attention mechanism as mentioned above. Along with this, we also use the $[CLS] \in R^d$ token from input-text feature representation of RoBERTa, and concatenated them forming the fusion vector as 
\begin{equation}
    H_{fusion} = concat(Q_{film},[CLS], Q_{att})
\end{equation}

where $H_{fusion} \in R^{1024+2 \times d}$. We pass this fusion vector through a fully connected layer followed by the sigmoid function for classification. So, the final output $\hat{y}$ would be:

\begin{equation}
    \hat{y} = Sigmoid(W^TH_{fusion} + b) 
\end{equation}

where $W \in R^{1024+2 \times d}$ and $b$ is a scalar and trainable parameters.  

\section{Experiments}

\subsection{The Dataset}\label{dataset}
We use the publicly available multimodal Twitter dataset collected by \cite{cai2019multi}. This dataset consists of 24k samples of the tweet, image, and image attributes. Further analysis of the dataset is shown in Table 1. This dataset is divided by \cite{cai2019multi} into the training set, validation set, and test set in the ratio 80\%:10\%:10\%, and we use the same split for a fair comparison. The dataset is preprocessed to separate words, emoticons, and hashtags with the NLTK toolkit. We present results on only one dataset consisting of text and image combined with image attribute since this is the only dataset where \emph{image attributes have been manually verified}. Original images are resized to 224, followed by center crop and normalization. We use data augmentation during training, including random center crop, random change of brightness, contrast, and image saturation. The text data is preprocessed to exclude the emoji information.

\begin{table}[h]
\begin{center}
\vspace{-8mm}
    \caption{Details of Twitter Dataset}
    \begin{tabular}{ccccc}
    \toprule
                   & \quad sentences\quad  & \quad positive\quad  & \quad negative\quad   &  \quad \% positive\quad    \\ \toprule
        Training    &  19816    & 8642      & 11174     & 43.62        \\ 
        Development &  2410     & 959       & 1451      & 39.76        \\ 
        Test        &  2409     & 959       & 1450      & 39.80        \\ \toprule
    \end{tabular}
    \vspace{-8mm}
\end{center}
\end{table}

\subsection{Baselines}
The baseline models for our experiment are as follows:

\begin{itemize}

\item \textbf{ResNet}: It is an image only model \cite{he2016deep} which is fine-tuned on the same Twitter multimodal dataset. 

\item \textbf{CNN}: A popular text only CNN \cite{DBLP:journals/corr/Kim14f} model which performs well on text classification problems. 

\item \textbf{Multi-dimensional Intra-Attention Recurrent Network (MIARN)}: \cite{tay-etal-2018-reasoning} proposed a novel architecture for text-only sarcasm detection by using 2D-attention mechanism to model intra-sentence relationships. 
 
\item \textbf{Hierarchical Fusion Model (HFM)}: Proposed by \cite{cai2019multi}, it takes text, image, and attribute feature as modalities. Features of the modalities are then reconstructed and fused for prediction. This is the only model besides ours that uses image attributes as an additional modality.

\item \textbf{D\&R Net}: \cite{xu-etal-2020-reasoning} preprocesses the image and text to form adjective-noun pairs (ANP). Then, they use a Decomposition and Relation Network (D\&R Net) to model cross-modality contrast using ANP and semantic association between the image and text.

\item \textbf{Res-bert}: \cite{pan2020modeling} implements Res-bert as a model to concatenate the output of image features from ResNet and text features from BERT. Since this model closely resembles our approach, it is an important baseline. 

\item \textbf{Intra and Inter-modality Incongruity (IIMI-MMSD)}: \cite{pan2020modeling} proposes  a BERT-based model, which concentrates on both intra and inter-modality incongruity for multimodal sarcasm detection. They use self-attention and co-attention mechanism to capture inter and intra-modality incongruity, respectively.

\item \textbf{Bridge-RoBERTa}: It is proposed by \cite{wang2020building}. The authors have used pre-trained RoBERTa and ResNet, and connected their vector spaces using a Bridge Layer. Further, to extract the relationship between text and image, they have used 2D-Intra-Attention layer.
\end{itemize}

\subsection{Experimental Settings and Hyper-Parameters}

The details of our experimental setup and hyper-parameters are as follows. We use pre-trained RoBERTa-base \cite{liu2019roberta} with 12 layers, and pre-trained ResNet-50 \cite{he2016deep} with 50 layers. For text and image attribute representation, we experiment with different number of layers of RoBERTa and find that 1 layer of encoder gives the best performance. So the comparison with baselines uses only 1 layer of RoBERTa encoder. We show the performance with different number of layers in the ablation studies. The model is run on NVIDIA Tesla V100-PCIE GPU. We use PyTorch 1.7.1 and Transformers 4.3.2 to implement our model. For evaluation we use F1-score, precision, recall, and accuracy as implemented in Scikit-learn.  We take Adam as our optimizer and set the learning rate for FiLMed network as 3e-4, for RoBERTa as 1e-6, and 1e-4 for co-attention layer. The batch size used is 32 for training. We also add weight decay of 1e-2 and gradient clipping set to 1.0. The maximum length of tokenised text is 360. We also take the standard dropout rate of 0.1. The model is fine-tuned for 15 epochs, and the model with the best F1-score on the validation set is saved and tested.

\subsection{Results and Discussion}

Table \ref{compare} shows the comparison of our proposed model with other baseline models. Our model outperforms the current state-of-the-art model \cite{wang2020building} on all the four metrics viz. F1-score, Precision, Recall, and Accuracy. Specifically, our model gives an improvement of 6.14\% on F1-score and 5.15\% on accuracy over the current SOTA from Bridge-RoBERTa model, thus verifying the effectiveness of our model. 
\begin{table}
\vspace{-5mm}
    \centering
        \caption{Comparison of baselines with our proposed model}
    \begin{tabular}{cccccc}
        \toprule
        Modality & Method & F1-score & Precision & Recall & Accuracy \\ \toprule 
        Image & ResNet \cite{he2016deep} & 0.6513 & 0.5441 & 0.7080 & 0.6476\\ \hline
        Text & CNN \cite{DBLP:journals/corr/Kim14f} & 0.7532 & 0.7429 & 0.7639 & 0.8003 \\
        & MIARN \cite{tay-etal-2018-reasoning} & 0.7736 & 0.7967 & 0.7518 & 0.8248 \\ \hline
        Image + Text & HFM \cite{cai2019multi} & 0.8018 & 0.7657 & 0.8415 & 0.8344\\
        & D\&R Net \cite{xu-etal-2020-reasoning} & 0.8060 & 0.7797 & 0.8342 & 0.8402\\
        & Res-Bert \cite{pan2020modeling} & 0.8157 & 0.7887 & 0.8446 & 0.8480\\
        & IIMI-MMSD \cite{pan2020modeling} & 0.8292 & 0.8087 & 0.8508 & 0.8605\\
        & Bridge-RoBERTa \cite{wang2020building} & 0.8605 & 0.8295 & 0.8939 & 0.8851\\ \hline
        & Our Method & \textbf{0.9219} & \textbf{0.9056} & \textbf{0.9387} & \textbf{0.9366} \\
        \toprule
    \end{tabular}
    \label{compare}
    \vspace{-5mm}
\end{table}

We can also verify from Table \ref{compare} that treating images or text independently does not perform well on the sarcasm detection problem. Intuitively, image-only models perform worse than text-only models as an image alone does not contain sufficient information to identify underlying sarcasm. Moreover, we can see that these unimodal approaches do not perform well for identifying sarcasm, and thus multimodal approaches are more suitable for sarcasm detection.  Further, from the Table \ref{compare}, we see that improvement in precision from baselines is more than the improvement in recall (the last two rows).  It shows that our approach can capture sarcastic tweets(+ve class) more accurately.

Our proposed model achieves better results than other multimodal approaches as it can capture the contradiction between the text and images in two stages: First, using FiLM, we get a representation of the image conditioned on the input text, thereby extracting image features that are incongruous to the text features. FiLM layers enable the GRU network over the input text to impact the neural network computation (ResNet in our case). This method helps to adaptively alter the ResNet network with respect to the input text, thereby allowing our model to capture inter modality incongruity between input text and image features. Second, the co-attention mechanism enables the image attributes to attend to each word in the input text. This helps us get a representation of high-level image features conditioned on the text. Thus, we can capture the inter-modality incongruity between input text and image attributes. Finally, since we have a representation of the image and image attributes, we need to include a representation of the input text to detect the sarcasm in a better way. Motivated by the approach taken in \cite{wang2020building}, using the output of $[CLS]$ token from RoBERTa for the final concatenation layer helps the proposed model to identify the underlying sarcasm using input-text, image and image attributes representation. Thus, conditioning the image and image attributes on input text using FiLM and co-attention respectively are effective for sarcasm detection.

\subsection{Model Analysis}
\begin{figure}
\centering
\mbox{\includegraphics[width=2.75cm, height=3.5cm]{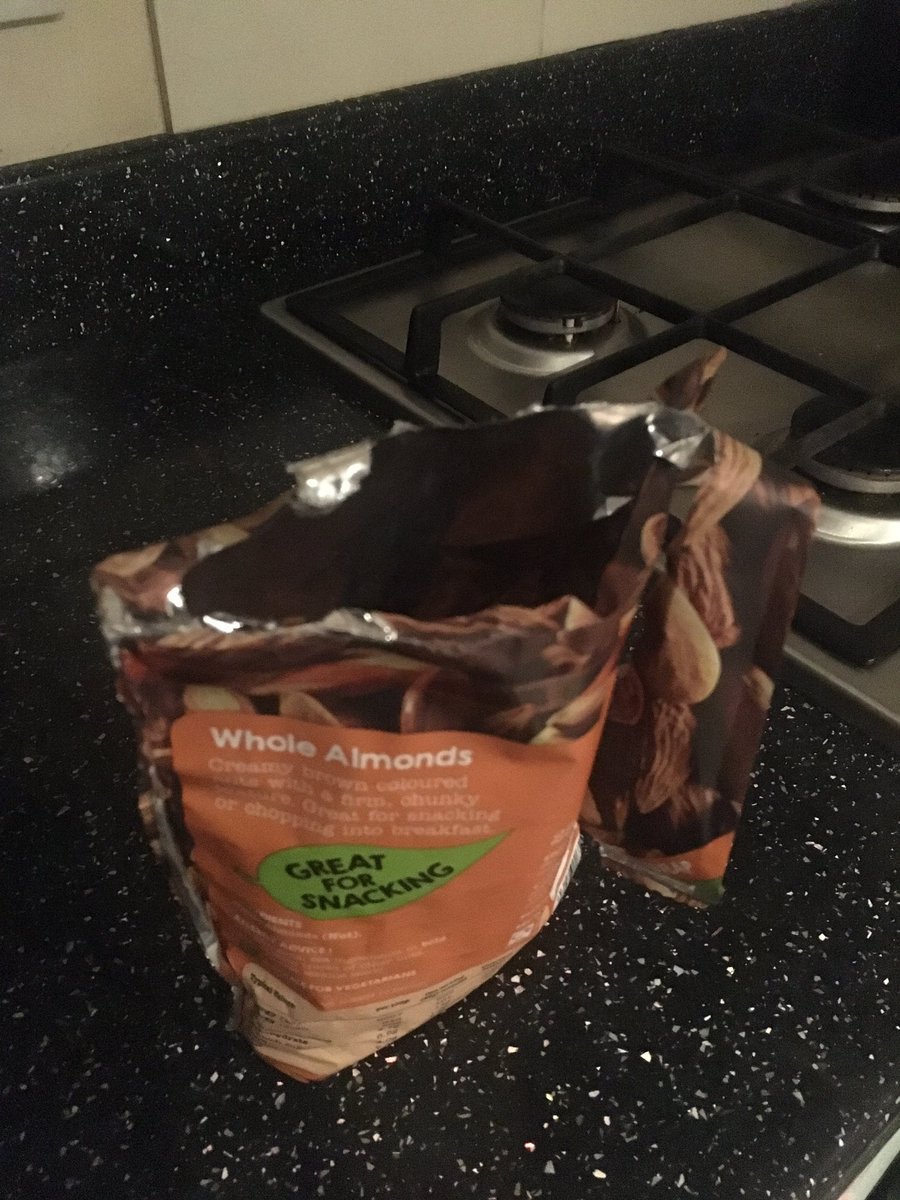}}
\mbox{\includegraphics[width=2.75cm, height=3.5cm]{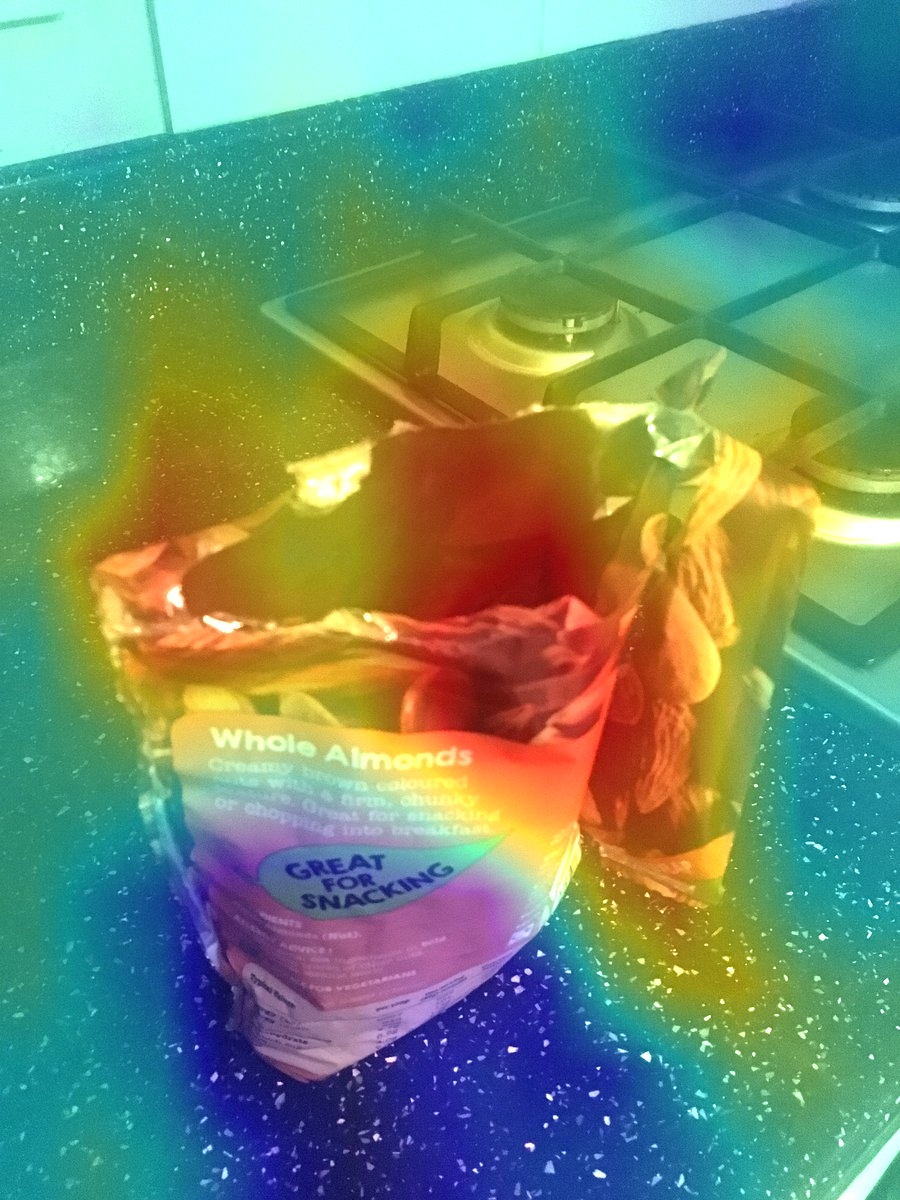}}
\hspace{5px}
 \mbox{\includegraphics[width=2.75cm, height=3.5cm]{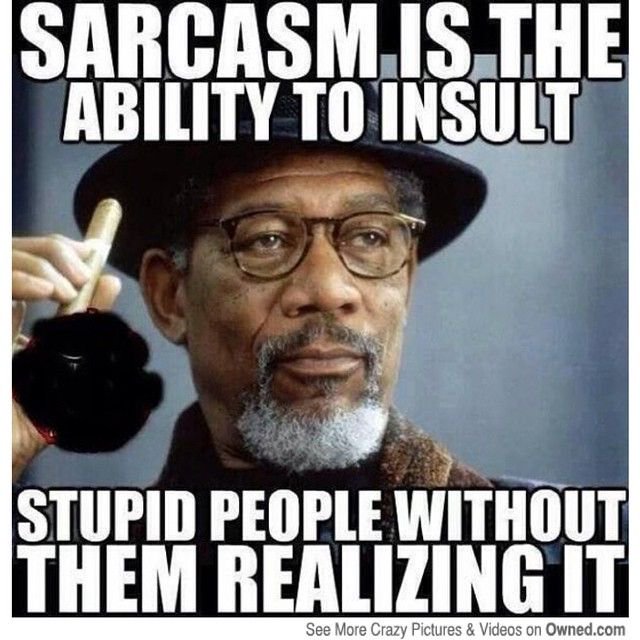}}
 \mbox{\includegraphics[width=2.75cm, height=3.5cm]{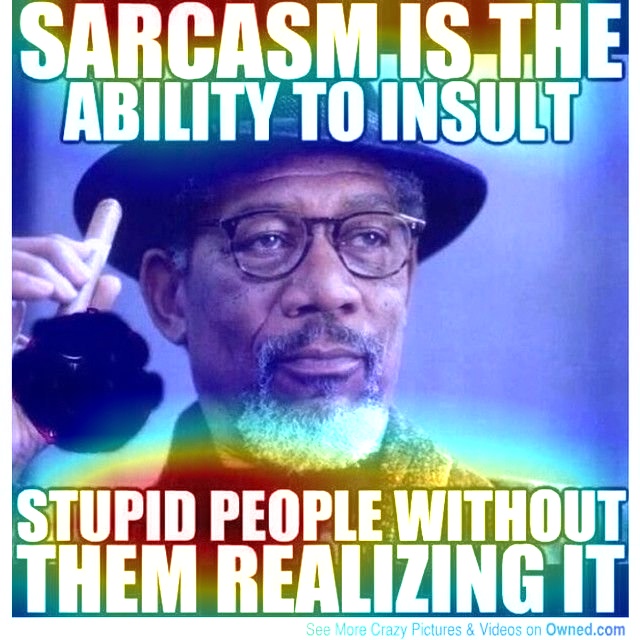}}
 \hspace{5px}\\
$(a)$ thanks god i have such a wonderful \quad \quad 
$(b)$ yup lol. \quad \quad \quad\quad\quad\quad\quad\quad\quad\quad\quad \quad \quad \quad \\
reseal in my pack of almonds \quad \quad \quad \quad \quad \quad \quad \quad \quad \quad \quad \quad \quad \quad \quad \quad \quad \quad \quad \quad \quad \quad 


 
\caption{Examples from the Twitter data showing attention visualization of sarcastic tweets.}
\label{attention}
\vspace{-8mm}
\end{figure}

To understand the visual significance of our model, we plot image attention representations. Fig. \ref{attention} illustrates that our model can attend to regions of image that are incongruous to text. The images show that regions corresponding to input text features are highlighted and will have larger activation than other regions of the image. Since final predictions are made using features of the highlighted areas, this operation gives an overall boost to the model performance. Moreover, we can verify from Figure \ref{attention}$(a)$ and \ref{attention}$(b)$ that our model is able to attend to text like "pack of almonds" and "stupid people" without the need to explicitly use noisy Optical Character Recognition (OCR), an approach used in previous works \cite{pan2020modeling,sangwan2020didn}.



\subsection{Ablation Study}

To evaluate the effectiveness of each component in our network, we conduct a detailed ablation study on the proposed architecture. Firstly, we remove the FiLM network which is represented as w/o FiLM. Secondly, co-attention between the visual attribute and the text is clipped, and the model is called w/o co-attention. Further, the importance of the $[CLS]$ token during the fusion is denoted by w/o cls. We experiment our approach with two other transformer models namely BERT and ELECTRA. We replaced RoBERTa with BERT and ELECTRA models and the resultant networks are called FiLM-Bert and FiLM-Electra respectively. The ablation results are shown in Table \ref{ablation}.

\begin{table}[h]
    \vspace{-8mm}
    \centering
    \caption{Results of ablation Studies. The `w/o' means removal of the component.}
    \begin{tabular}{ccccc}
    \toprule
         Ablation & F1-score & Precision & Recall & Accuracy \\ \toprule
         w/o FiLM &  0.6217 & 0.5667 & 0.6890 & 0.6660 \\
         w/o co-attention & 0.7607 & 0.7360 & 0.7871 & 0.8029 \\ 
         w/o cls & 0.7638 & 0.7225 & 0.8104 & 0.8003 \\
         FiLM-Electra & 0.7683 & 0.7178 & 0.8265 & 0.8013 \\
         FiLM-Bert & 0.7727 & 0.7131 & 0.8439 & 0.8026 \\
         Our Method & $\textbf{0.9219}$ & \textbf{0.9056} & \textbf{0.9387} & \textbf{0.9366} \\
         \toprule
    \end{tabular}
   \label{ablation}
   
    \vspace{-5mm}
    
\end{table}
   We can see that the removal of FiLM (w/o FiLM) significantly hampers the model's performance. Next, we eliminate the co-attention module (w/o co-attention) and concatenate the output from FiLM and RoBERTa. This decreases the model performance which implies that capturing incongruity between image and textual features through co-attention is important for sarcasm detection. Further, output from the [CLS] token positively contributes to the model as removing the [CLS] token (w/o cls) reduces the metric scores. When we try our network with the BERT and ELECTRA model (FiLM-Bert and FiLM-Electra), then we observe sharp decline in the performance. This shows that RoBERTa is better at harnessing textual features. The use of larger training data and batch size in RoBERTa gives it an edge over the original BERT model. From the above results, we can conclude that FiLM network fused with RoBERTa and attention mechanism helps to capture incongruity between image and text modality, thereby effectively learning the underlying sarcasm. 
   \\
 
\begin{table}[h]
    \vspace{-8mm}
    \centering
    \caption{Ablation study with different layers of RoBERTa model.}
    \begin{tabular}{cccccc}
    \toprule
         Layers & F1-score & Precision & Recall & Accuracy & Training Time (per epoch) \\ \toprule
         12 & 0.8365 & 0.8242 & 0.8492 & 0.8679 & 15.45 minutes \\ 
        5 & 0.8434 & 0.8217 & 0.8665 & 0.8711 & 8.96 minutes \\
         2 & 0.9215 & 0.9043 & 0.9393 & 0.9363 & 6.02 minutes  \\
         1 & $\textbf{0.9219}$ & \textbf{0.9056} & \textbf{0.9387} & \textbf{0.9366} & \textbf{5.10 minutes} \\
         \toprule
    \end{tabular}
   \label{ablation-layers}
   
    \vspace{-5mm}
    
\end{table}
  Motivated by \cite{rogers2020primer}, we study the layer transferability in the RoBERTa model for our task. Instead of directly taking the output from the $12^{th}$ attention encoder layer, we consider the output from different layers to represent the textual features $P$. The rest of the model architecture is unaltered. Results are presented in Table \ref{ablation-layers}. The authors of \cite{rogers2020primer} proposed that middle layers of the BERT model are most prominent in representing syntactic information. We observe a similar trend in Table \ref{ablation-layers}. We get best results when we use the output of the first encoder layer. Our findings imply that initial layers of the RoBERTa model are better at encoding the syntactic representation and have most information about the linear word order. The final layers are more task-specific and give better performance in applications where we simply attach a classifier on top of the transformer model for downstream tasks. Using only one layer also helps in reducing the model size and the training time as seen in the last column of Table \ref{ablation-layers}.
\section{Conclusion and Future Work}
The present work tackles the problem of sarcasm detection through capturing inter-modality incongruity. The proposed architecture handles the inter-modality incongruity in two ways: the first uses the co-attention, and the second is via FiLM network. Comparison with baselines on Twitter benchmark datasets reveals that the proposed architecture can better capture the contradiction present between the image and text modality. The ablation study highlights the importance of FiLM and co-attention layer between the image, image attribute embeddings and the text embeddings. Comparison with several baselines on benchmark dataset shows the effectiveness  and superiority of our model.

\bibliographystyle{splncs04}
\bibliography{sample-base}


\end{document}